\documentclass[aps,showpacs,preprintnumbers,amsmath,amssymb,showkeys]{revtex4}

\usepackage{graphicx}
\usepackage{color}

\def \be  {\begin{equation}}
\def \ee  {\end{equation}}
\def \ee  {\end{equation}}
\def \bea {\begin{eqnarray}}
\def \eea {\end{eqnarray}}

\newcommand{\nn}{\nonumber}

\begin{document}

\preprint{ECTP-2013-04}

\title{Corrections to entropy and thermodynamics of charged black hole using generalized uncertainty principle}

\author{Abdel Nasser Tawfik}
\email{a.tawfik@eng.mti.edu.eg}
\affiliation{Egyptian Center for Theoretical Physics (ECTP), Modern University for Technology and Information (MTI), 11571 Cairo, Egypt}
\affiliation{World Laboratory for Cosmology And Particle Physics (WLCAPP), Cairo, Egypt}

\author{Eiman Abou El Dahab}
\affiliation{Faculty of Computer Sciences, Modern University for Technology and Information (MTI), 11571 Cairo, Egypt}
\affiliation{World Laboratory for Cosmology And Particle Physics (WLCAPP), Cairo, Egypt}

\date{\today}

\begin{abstract}

Recently, there has been much attention devoted to resolving the quantum corrections to the Bekenstein-Hawking (black hole) entropy, which relates the entropy to the cross-sectional area of the black hole horizon. Using generalized uncertainty principle (GUP), corrections to the geometric entropy and thermodynamics of black hole will be introduced. The impact of GUP on the entropy near the horizon of three types of black holes; Schwarzschild, Garfinkle-Horowitz-Strominger and Reissner-Nordstr\"om is determined. It is found that the logarithmic divergence in the entropy-area relation turns to be positive. The entropy $S$, which is assumed to be related to horizon's two-dimensional area, gets an additional terms, for instance $2\, \sqrt{\pi}\, \alpha\, \sqrt{S}$, where $\alpha$ is the GUP parameter.

\end{abstract}. 


\maketitle


\section{Introduction}

The finding that the black holes should have well-defined entropy and temperature represented one of the greatest achievements in recent astrophysics \cite{hist1a,hist1b,hist1c}. In statistical physics and thermodynamics, the entropy relates the number of macrostates to that of microstates of the system of interest. Furthermore, the entropy can be defined as the amount of additional information needed to specify the exact physical state of a system. In general relativity, the entropy of black hole got a novel definition. It is a pure geometric quantity so that when comparing black hole with a thermodynamic system, we find an important difference. Whether the black hole has interior  degrees of freedom corresponding to its entropy, the Bekenstein-Hawking entropy delivered an answer to this and characterized the statistical meaning \cite{hist1a,hist1b,hist1c}. The horizon's two-dimensional area defines the entropy. For completeness, we mention this entropy is defined by Boltzmann-Gibbs statistics. The Tsallis entropy, which is a generalization of Boltzmann-Gibbs entropy, would have another different dependence \cite{tsallis}. Identification and/or counting of the microstates was highlighted in Ref. \cite{refff2}. A non-vanishing entropy of extremal charged black holes was introduced \cite{Wotae}. It was suggested that this presumably lies within the framework of the quantum gravity. For example, the string theory \cite{hist4} and loop quantum gravity \cite{hist5} succeeded in presenting an statistical explanation formulated in an entropy-area law. The proportionality relating black hole entropy with area was derived from classical thermodynamics, as well \cite{sAclass}. In this regard, we recall that a black hole is nothing but a classical solution in general theory of relativity with some special properties.

As discussed in \cite{refff2,Tawfik:2015pqa}, the quantum correction of geometric entropy of charged black hole would avoid being biased in favour of a certain theory of quantum gravity.  Furthermore, the tree-level quantum correction \cite{hist7} has been verified in different studies \cite{hist8}. The correction to the Bekenstein-Hawking entropy, which relates the entropy to the cross-sectional area of the black hole horizon, has a series of terms  \cite{refff2,Tawfik:2015pqa}. The coefficient of the leading-order correction, the logarithmic term, is suggested as a discriminator of prospective fundamental theories for quantum gravity \cite{hist9}. It is essential to suggest a method that fixes it, but it should not depend on the utilized models for the quantum gravity. For instance, this might be the holographic principle \cite{hist16}. 

It is found that the covariant entropy bound in the Friedmann-Lemaitre-Robertson-Walker (FLRW) Universe gives an indication to the holographic nature in terms of temperature and entropy \cite{everlind1}. The cosmological boundary can be chosen as the cosmological apparent horizon instead of the event horizon of a black hole. In light of this, it was argued that the statistical (informational) entropy of a black hole can be calculated using the brick wall method (BWM) \cite{entr5}. In order to avoid the divergence near the event horizon, a cutoff parameter would be utilized. Since the degrees of freedom would be dominant near the horizon, the brick wall method could be replaced by a thin-layer model making the calculation of entropy possible \cite{entr6a,entr6b,entr6c,entr6d,entr6e,entr6f,entr6g,entr6h}. The entropy of FLRW Universe is given by time-dependent metric. 

In brick wall model, the usual position-momentum uncertainty relation is given  by
\bea
\Delta x\; \Delta p \geq \frac{\hbar}{2}.
\eea
Then, the entropy can be calculated as follows.
\bea
S_0 &=& \beta^2 \left.\frac{\partial F_0}{\partial \beta}\right|_{\beta=\beta_H}
= \frac{\beta^2}{\pi} \int_{r_++\epsilon}^{L} dr\; \frac{1}{\sqrt{f}} \int_{m \sqrt{f}}^{\infty} d \omega \left.\frac{\omega e^{\beta \omega} \left(\frac{\omega^2}{f} - m^2\right)^{1/2}}{\left(e^{\beta \omega}-1\right)^{2}} \right|_{\beta=\beta_H},
\eea
where $\beta$ is the inverse temperature, $F_0$ is the free energy and $L$ and $\epsilon$ are infrared and ultraviolet regulators, respectively. $\beta_H$ is the inverse Hawking temperature. In massless limit ($m=0$), the entropy reads 
\bea
S_0^{non-ext} &\approx & \frac{1}{12}\, \ln \left(\frac{1}{2\, \Lambda\; \epsilon}\right), 
\eea
where $\Lambda$ stands for the cosmological constant. In deriving this expression the infrared divergent $L$-term is entirely cancelled.
 
For the external case, we can assume that $\beta\rightarrow\infty$.
\bea
S_0^{ext} &=& \beta^2\, \left. \frac{\partial F_0}{\partial \beta}\right|_{\beta\rightarrow\infty} = 0.
\eea
Opposite to the extremal case, we assume a zero-temperature quantum mechanical system around the black hole. Then, the entropy reads
\bea
S_0^{ext} &\approx& \ln \left(\frac{1}{2\, \Lambda\, \epsilon}\right),
\eea
which can be interpreted as the physical limit that $\Lambda$ should be less than $1/(2\epsilon)$.

Although these results include a logarithmic divergence, they will be changed due to the GUP effects that will be outlined in the section that follows  \cite{sabine}. Section \ref{sec:corrctTherm} introduces corrections to the black hole thermodynamics using GUP approach. The corrections to black hole entropy due by GUP are studied in section \ref{sec:dltasgup}. The GUP approach is applied to three types of black holes; Schwarzschild,  Garfinkle-Horowitz-Strominger and Reissner-Nordstr\"om. The conclusions are outlined in section \ref{sec:co}.

\section{Thermodynamics near event horizon and GUP approach}
\label{sec:corrctTherm}

The GUP approach introduced in \cite{alii} claims to make predictions for maximum observable momentum and minimal measurable length. Accordingly, $[x_i, x_j]=[p_i, p_j]=0$ can be produced via the Jacobi identity, and therefore 
\bea
[x_i, p_j]\hspace{-1ex} &=&\hspace{-1ex} i \hbar\hspace{-0.5ex} \left[  \delta_{ij}\hspace{-0.5ex} - \hspace{-0.5ex} \alpha \hspace{-0.5ex}  \left( p\, \delta_{ij} +
\frac{p_i p_j}{p} \right) + \alpha^2 \hspace{-0.5ex} \left( p^2\, \delta_{ij}  + 3\, p_{i}\, p_{j} \right) \hspace{-0.5ex} \right], \label{eq:alfaa}
\label{comm01}
\eea
where the parameter $\alpha = {\alpha_0}/{M_{p}c} = {\alpha_0 \ell_{p}}/{\hbar}$ and $M_{p} c^2$ stands for Planck's energy. $M_{p}$ and $\ell_{p}$ is Planck's mass and length, respectively. Apparently, Eq. (\ref{comm01}) implies the existence of a minimum measurable length and a maximum measurable momentum
\bea
\Delta x_{min} & \approx & \alpha_0\ell_{p} \label{dxmin} , \\
\Delta p_{max} &\approx & \frac{M_{p}c}{\alpha_0} \label{dpmax},
\eea
where the uncertainties $\Delta x \geq \Delta x_{min}$ and $\Delta p \leq \Delta p_{max}$.
Accordingly, for a particle having a distant origin and an energy scale comparable to the planckian one, the momentum would be a subject of a  modification 
\bea
p_i &=& p_{0i} \left(1 - \alpha p_0 + 2\alpha^2 p_0^2 \right), \label{mom1} \\
 x_i &=&  x_{0i},
\eea
where $p_0^2=\sum_i p_{0i} p_{0i}$ and  $p_{0i}$ are the components of the low energy momentum. The operators $p_{0j}$ and $ x_{0i}$ satisfy the canonical commutation relation $[x_{0i}, p_{0j}] = i \hbar~\delta_{ij}$. Having the standard representation in position space, then  $p_{0i} = -i \hbar \partial/\partial{x_{0i}}$ and $x_{0i}$ would represent
the spatial coordinates operator at low energy \cite{alii}. Then the uncertainty relation in natural units reads
\bea
\Delta x\, \Delta p \geq \frac{\hbar}{2} \left[1-2\, \alpha\, \langle p\rangle + 4\, \alpha^2\, \langle p^2\rangle \right],
\eea
the volume of phase cell in the $1+1$ dimension is changed from $2\, \pi$ to 
\bea
2\pi\, \left(1 - 2\, \alpha\; p + 4\, \alpha^2\; p^2\right),
\eea
and the number of quantum states with energy less than $\epsilon$ \cite{reffff1}
\bea
n_0(\omega) &=& \frac{1}{2 \pi} \int dr \; d p_r
= \frac{1}{\pi} \int^L_{r_+\, + \epsilon} d r \, \frac{1}{\sqrt{f}}\; \left(\frac{\omega^2}{f}-m^2\right)^{1/2}, \label{eq:n11}
\eea
where $m$ in the mass of the scalar field and $\omega$ is a parameter of the substitution of Klein-Gordon equation. The expression given in Eq. (\ref{eq:n11}) will be changed to
\bea
n_I(\omega) &=& \frac{1}{2 \pi} \int dr \; d p_r\; \frac{1}{1-2\, \alpha\, p + 4\, \alpha^2\,  p^2}, \label{eq:nz1} \nonumber \\
                    &=& \frac{1}{2 \pi} \int dr \; \frac{1}{\sqrt{f}} \frac{\left(\frac{\omega^2}{f}-m^2\right)^{1/2}}{1-2\, \alpha\, \left(\frac{\omega^2}{f}-m^2\right)^{1/2} + 4\, \alpha^2\,  \left(\frac{\omega^2}{f}-m^2\right)}, \label{eq:nz2}
\eea
where $r$ and $f$ are given in Eqs. (\ref{eq:f1}) and (\ref{eq:r1}), respectively. The WKP approximation (see Eqs. (\ref{eq:pr2}) and (\ref{eq:p2})) is implemented.

At Hawking temperature,  Eq. (\ref{eq:nz1}) can be used to derive the free energy  \cite{reffff1} 
\bea \label{eq:fe1}
F_0 &=& - \frac{1}{\pi} \int^L_{r_++\epsilon}\; dr \frac{1}{\sqrt{f}} \; \int_{m \sqrt{f}}^{\infty} \; d\omega \frac{\left(\frac{\omega^2}{f}-m^2\right)^{1/2}}{e^{\beta \omega}-1},
\eea
which turns to be a subject of a change to
\bea \label{eq:fe2}
F_I &=& -\int_{m \sqrt{f}}^{\infty} d \omega \frac{n_I(\omega)}{e^{\beta \omega}-1}, \nonumber \\
    &=& -\frac{1}{\pi} \int dr \frac{1}{\sqrt{f}} \int_{m \sqrt{f}}^{\infty} d \omega \frac{\left(\frac{\omega^2}{f}-m^2\right)^{1/2}}{\left(e^{\beta \omega}-1\right) \left[1-2 \alpha \left(\frac{\omega^2}{f}-m^2\right)^{1/2} + 4 \alpha^2 \left(\frac{\omega^2}{f}-m^2\right)\right]}.
\eea

\section{Black hole entropy and GUP approach}
\label{sec:dltasgup}

Near the event horizon i.e., in the range $(r_+, r_++\epsilon)$, $f\rightarrow 0$, the entropy can be deduced from Eq. (\ref{eq:fe2})  
\bea
S_0 &=& \frac{\beta^2}{\pi} \int^{L}_{r_++\epsilon} d r \frac{1}{\sqrt{f}} \int^{\infty}_{m \sqrt{f}} d\omega \left.\frac{\omega e^{\beta \omega} \left(\frac{\omega^2}{f}-m^2\right)^{1/2}}{\left(e^{\beta \omega}-1\right)^2}\right|_{\beta=\beta_H}. \label{eq:S01}
\eea
Once again, the entropy given in Eq. (\ref{eq:S01}) will be changed to
\bea
S_I &=& \frac{\beta^2}{\pi} \int dr \frac{1}{\sqrt{f}} \int_{m \sqrt{f}}^{\infty} \frac{\omega \left(\frac{\omega^2}{f}-m^2\right)^{1/2} e^{\beta \omega}}{e^{2\beta \omega - 2} \left[1-2 \alpha \left(\frac{\omega^2}{f}-m^2\right)^{1/2} + 4 \alpha^2 \left(\frac{\omega^2}{f}-m^2\right)\right]} d\omega, \nonumber \\
&=& \frac{1}{\pi} \int_{r_+}^{r_++\epsilon} dr \frac{1}{\sqrt{f}} \int_{0}^{\infty} \frac{f^{-1/2}\, \beta^{-1}\, x^2}{(1-e^{-x})(e^x-1) \left[1-2 \alpha \frac{x}{\beta \sqrt{f}} + 4 \alpha^2 \frac{x^2}{\beta^2 f}\right]} d x,
\eea
where $x=\beta\, \omega$. We note that as $f\rightarrow 0$, then $\omega^2/f$ is the dominant term in the bracket containing $\omega^2/f - m^2$. We are interested in the thermodynamic contributions just near the horizon
$r_+,r_++\epsilon$, which corresponds to a proper distance of the order of the minimal length, which can be related to $\alpha$ using natural units ($\alpha={\alpha_0 \ell_{p}}/{\hbar}=2 \pi \ell_{p}\approx \ell_{p}$). So we have from Eq. (\ref{eq:ds2})
\bea
\alpha &=& \int_{r_+}^{r_++\epsilon} \frac{d r}{\sqrt{f(r)}},
\eea
which apparently sets a lower bound to $\alpha$. 
Then the entropy reads
\bea
S_I &=& \frac{1}{\pi\; \alpha}  \int_{r_+}^{r_++\epsilon} \frac{d r}{\sqrt{f(r)}} \int_0^{\infty} d X\, \frac{a^2\, X^2}{\left(e^{\frac{a\, X}{2}} - e^{-\frac{a\, X}{2}}\right)^2\; \left(1-2\, X + 4\, X^2\right)},
\eea
where 
\bea
x &=& \frac{\beta}{\alpha}\, \sqrt{f}\, X = a\, X.
\eea
Then
\bea
S_I &=& \frac{1}{\pi}\; \Sigma_I = \frac{1}{\pi}\; \int_0^{\infty}\, \frac{a^2\, X^2}{\left(e^{\frac{a\, X}{2}} - e^{-\frac{a\, X}{2}}\right)^2\; \left(1-2\, X + 4\, X^2\right)}\, d\, X.
\eea
We note  that as $r\rightarrow r_+$, $f\rightarrow 0$, then $a\rightarrow 0$ and
\bea
\lim_{a\rightarrow 0}  \frac{a^2 \, X^2}{\left(e^{a \, X/2}-e^{-a \, X/2}\right)^2} =1.
\eea
Therefore, 
\bea
\Sigma_I &=& \int_0^{\infty} \frac{d\, X}{1-2 \, X + 4 \, X^2} = \frac{2\, \pi}{3 \sqrt{3}},
\eea
and
\bea
S_I &=& \frac{1}{\pi} \; \Sigma_I = \frac{2}{3 \, \sqrt{3}}.
\eea
So far, we conclude that $S_I$ is finite. It does not depend on any parameter.

We note that in contrast to the case of brick wall method, there is no divergence within the just vicinity near the horizon due to the effect of the generalized uncertainty relation on the quantum states.

In the sections that follows, we estimate the corrections in the geometric entropy of three types of black holes; Schwarzschild,  Garfinkle-Horowitz-Strominger and Reissner-Nordstr\"om  due to the GUP approach.

\subsection{Schwarzschild black hole}

In the Schwarzschild  gauge, the metric and field tensors, respectively,  are conjectured to be expressed as
\bea
d s^2 &=& - f(r) d t^2 + \frac{1}{f(r)} d r^2, \label{eq:ds2}\\
F_{rt} &=& F_{rt} (r). \label{eq:frt}
\eea
The function $f(r)$ in the static solution is defined as 
\bea
f(r) &=& 1- \frac{M}{\Lambda} e^{-2 \Lambda r} + \frac{Q^2}{4 \Lambda^2} e^{-4 \Lambda r}, \label{eq:f1}
\eea
where $M$ is the mass of black hole and $Q$ gives its charge. The outer event horizon has the radius
\bea
r_{+} &=& \frac{1}{2 \Lambda} \ln\left[\frac{M}{2 \Lambda} + \sqrt{\left(\frac{M}{2 \Lambda}\right)^2 - \left(\frac{Q}{2 \Lambda}\right)^2} \right]. \label{eq:r1}
\eea
In light of this, its derivative vanishes and the Klein-Gordon equation is reduced to
\bea
\frac{d^2 R}{d r^2} +\frac{1}{f} \frac{d f}{d r} \frac{d R}{d r} + \frac{1}{f} \left(\frac{\omega^2}{f} - m^2\right) R &=& 0,
\eea
where $\phi(r) = \exp\left(-i \omega t\right) R(r)$. Using WKB approximation, then $R \sim \exp\left(i S(r)\right)$, 
\bea \label{eq:pr2}
p_r^2 &=& \frac{1}{f} \left(\frac{\omega^2}{f} - m^2\right),
\eea 
and $p_r=d S/d r$ and 
\bea\label{eq:p2}
p^2 = \frac{\omega^2}{f} - m^2.
\eea

In natural units, $\hbar=c=G=k_B=1$, the line element in Schwarzschild black hole reads
\bea
d s^2 &=& -\left(1-2\, \frac{M}{r}\right) d\, t^2 + \left(1-2\, \frac{M}{r}\right)^{-1}\, d\, r^2 + r^2\, d\,\Omega_2^2.
\eea
Then, Hawking radiation temperature $T$, horizon area $A$ and entropy $S$, respectively, read 
\bea
T &=& \frac{1}{4 \pi r_H} = \frac{1}{8 \pi M},\\
A &=& 4 \pi r_H^2 = 16 \pi M^2,\\ 
S &=& \pi r_H^2 = 4 \pi M^2,
\eea
where $r_H = 2 M$ is the location of the black hole horizon. The increase (decrease) in the horizon area due to absorbing (radiating) a particle of energy $d\, M$ can be expressed as
\bea
d\, A &=& 8\, \pi\, r_H\, d r_H = 32\, \pi\, M\, d M.
\eea
This particle is conjectured to satisfy Heisenberg's uncertainty relation $\Delta\, x_i \, \Delta\, p_j \geq \delta_{ij}$.

There are different approaches for GUP. For review, reader are referred to \cite{sabine,Tawfik:2014zca}.
 
\subsubsection{Quadratic QUP approach}

When Heisenberg's uncertainty relation is replaced by quadratic GUP \cite{gupp1}
\bea \label{eq:dltGup}
\Delta\, x_i \geq \frac{1}{\Delta\, p_i} + \alpha^2 \, \Delta\, p_i,
\eea
where  Plank length $l^2_{pl}=((\hbar \, G_d)/c^3)^{1/2}$ equals unity in natural units. From Eq. (\ref{eq:dltGup}), we have
\bea
\Delta\, p_i &\geq& \frac{1}{\Delta\, x_i} \left[1+\frac{\alpha^2}{(\Delta\, x_i)^2} + 2 \left(\frac{\alpha^2}{(\Delta\, x_i)^2}\right)^2+\cdots \right].
\eea
Then,
\bea
d\, A &=& 32\, \pi\, M\, \frac{1}{\Delta \, x}, \label{eq:Aa1} \\
d\, A_{GUP} &=& d\, A \left[1+\frac{\alpha^2}{(\Delta\, x)^2} + 2 \left(\frac{\alpha^2}{(\Delta\, x)^2}\right)^2+\cdots \right]. \label{eq:Aa2}
\eea
If we utilize the notations of Ref. \cite{refff2}, then  
\bea
\Delta\, x &=& 2\, r_H = \sqrt{\frac{A}{\pi}}. 
\eea
After straightforward integration, we derive
\bea
A_{GUP} &=& A + \alpha^2\, \pi\, \ln A - 2 \left(\alpha^2\, \pi\right)^2\, \frac{1}{A} + \cdots. \label{eq:Aa2b}
\eea
The Bekenstein-Hawking area law states that $S=A/4$. Then, the black hole's entropy can be derived from Eq. (\ref{eq:Aa2b}) 
\bea
S_{GUP} &=& S + \frac{\alpha^2}{4}\, \pi \, \ln\, S - \left(\alpha^2\, \pi\right)^2\, \frac{1}{8\, S} - \cdots + C,
\eea
where $C$ is an arbitrary constant. We notice that the coefficient of the logarithmic correction term is positive. It is a leading-order correction known as logarithmic ''prefactor''.  This result obviously contradicts the one given in \cite{refff2}. 

\subsubsection{Linear GUP approach}

If we use GUP introduced in \cite{alii},
\bea
\Delta\, x \, \Delta\, p & \geq & \left[1-2\, \alpha \, \Delta\, p\right],
\eea 
then  
\bea
\Delta\, p & \geq & \frac{1}{\Delta \, x} \left(\frac{1}{1+\frac{2\, \alpha}{\Delta \, x}}\right).
\eea
Accordingly, the area and entropy, respectively,  can be re-written as
\bea
A_{GUP} &=& A - 4 \alpha\, \sqrt{\pi}\, \sqrt{S}+ 8\, \pi\, \alpha^2\, \ln\left(\sqrt{\frac{A}{\pi}}+2\,\alpha\right),\\
S_{GUP} &=& S - 2\, \alpha\, \sqrt{\pi}\, \sqrt{S} + \alpha^2\, \pi\, \ln S + C,
\eea where $\alpha\ll \sqrt{A/\pi}$. We notice that the coefficient of $\ln S$ is also positive, but the entropy gets an additional term,  $2\, \alpha\, \sqrt{\pi}\, \sqrt{S}$. $C$ is an arbitrary constant.

\subsection{Garfinkle-Horowitz-Strominger black hole}

The line element in Garfinkle-Horowitz-Strominger space-time \cite{refff4} reads
\bea
d\, s^2 &=& - \left(1-2\frac{M}{r}\right) d\, t^2 + \left(1-2\frac{M}{r}\right)^{-1} + r(r-2\, a) d\, \Omega_2^2,
\eea
where $a=Q^2/2M$ and $Q$ is the electric charge of the black hole. The horizon area and entropy, respectively, are
\bea
A &=& 4\, \pi\, r_H \left(r_H-2\, a\right) = 16\, \pi\, M \left(M-a\right), \\
S &=& 4\, \pi\, M \left(M-a\right).
\eea
Assuming that $a$ is constant, then the change in $A$ reads
\bea
d\, A &=& 16\, \pi\, M \left(2\, M-a\right)\, d M.
\eea
When assuming, as before, that $\Delta\, x=2\, r_H$ and $a\ll r_H$, then Eq. (\ref{eq:Aa2}) leads to corrected area and entropy
\bea
A_{GUP} &=& A+\alpha^2\, \pi\, \ln A + 8\, a^2\, \frac{\pi^2\, \alpha^2}{A} + 8\, a\, \frac{\pi^{3/2}\, \alpha^2}{A^{1/2}} - 2 \frac{\left(\alpha^2\, \pi\right)^2}{A} + 16\,a^2  \frac{\left(\alpha^2\, \pi\right)^2\, \pi}{A^2}, \\
S_{GUP} &=& S+\frac{\alpha^2\, \pi}{4}\, \ln S + a^2\, \frac{\pi^2\, \alpha^2}{2\, S} + a\, \frac{\pi^{3/2}\, \alpha^2}{S^{1/2}} - \frac{\left(\alpha^2\, \pi\right)^2}{8\, S} + a^2  \frac{\left(\alpha^2\, \pi\right)^2\, \pi}{4\, S^2}+C.
\eea

If we use GUP that introduced in \cite{alii},
\bea
A &=& 4\, \pi\, r_H^2 \left(1-\frac{2\,a}{r_H}\right), \label{eq:aaGUP}
\eea
which is valid for $a\ll r_H$. Expression (\ref{eq:aaGUP}) can be used to estimate $r_H$, 
\bea
r_H &=& \frac{1}{2}\sqrt{\frac{A}{\pi}} + a.
\eea
Then, the change in area of black hole is
\bea
d\, A_{GUP} &=& \left(1+\frac{2\, \alpha}{\Delta\, x}\right)^{-1}\, d\, A.
\eea
After integration, we get expressions for corrected area and entropy
\bea
A_{GUP} &=& A-4\, \alpha\, \sqrt{\pi}\, \sqrt{A} + 8\, \alpha\, \pi\, (a+\alpha)\, \ln\left(\sqrt{\frac{A}{\pi}}+2(a+\alpha)\right), \\
S_{GUP} &=& S-2\, \alpha\, \sqrt{\pi}\, \sqrt{S} + \alpha\, \pi\, (a+\alpha)\, \ln S + C,
\eea
where $C$ is an arbitrary constant. It is assumed that $a+\alpha\ll (A/\pi)^{1/2}$. Again, we notice that the logarithmic term is positive and an additional term $2\, \alpha\, \sqrt{\pi}\, \sqrt{S}$ appears.

\subsection{Reissner-Nordstr\"om Black Hole}

In space-time of Reissner-Nordstr\"om black hole, the line element  is given by \cite{RN1,RN2}
\bea
d\, s^2 &=& - \left(1-2\frac{M}{r}+\frac{Q^2}{r^2}\right) d\, t^2 + \left(1-2\frac{M}{r}+\frac{Q^2}{r^2}\right)^{-1} d\, r^2 + r^2 \, d\, \Omega_2^2,
\eea
where $r_{\pm} = M \pm (M^2-Q^2)^{1/2}$ are locations of outer and inner horizon, respectively.  The area and entropy of outer horizon, are given as $A = 4\, \pi\, r_+^2$ and $S=\pi\, r_+^2$, respectively. When the electric charge $Q$ is taken invariable and $Q\ll r_H$ \cite{refff5}, then the corrected area and entropy are
\bea
A_{GUP} &=& A + \alpha^2\, \pi\, \ln A 
- 16 \alpha^2\, \pi^2\, \left(\frac{Q^2}{A} + 8 \frac{Q^4}{A^2}\right) 
- 2 \left(\alpha^2\, \pi\right)^2 \left(\frac{1}{A}+16 \frac{Q^2}{A^2} + \frac{32\, Q^4}{3\, A^3} \right), \hspace*{10mm} \\
S_{GUP} &=& S + \frac{\alpha^2\, \pi}{4}\ln S -\alpha^2\, \pi^2\left(\frac{Q^2}{S}+2\frac{Q^4}{S^2}\right) - \left(\alpha^2\pi\right)^2 \left(\frac{1}{8\, S}+\frac{Q^2}{2\, S^2} + \frac{Q^4}{12\, S^3}\right) +C+\cdots.
\eea

If we use for GUP the approach which was introduced in \cite{alii}, then the corrected area is 
\bea
d\, A_{GUP} &=& \frac{A-4\, \pi Q^2}{A-4\, \pi Q^2+2\, \alpha\, \sqrt{\pi\, A}}\, d\, A,
\eea
which can be integrated
\bea
A_{GUP} &=& A - 4\alpha \sqrt{\pi} \sqrt{A} + 4 \alpha^2 \pi \ln\left(\sqrt{A}\left(\sqrt{A}+2 \alpha \sqrt{\pi}\right)-4\,\pi\,Q^2\right) \nonumber \\
&-& 8\alpha \pi\sqrt{\pi}\left(\alpha^2+2\,\pi\,Q^2\right)\left[\frac{1}{\kappa} \ln \frac{-\sqrt{A}+2\alpha\sqrt{\pi}-\kappa}{-\sqrt{A}+2\alpha\sqrt{\pi}+\kappa}\right],
\eea
where $\kappa=\sqrt{4 \alpha^2 \pi + 16 \pi Q^2}$. For finite $\alpha$ and $Q\ll \sqrt{A}$, the entropy reads
\bea
S_{GUP} &=& S - 2 \alpha \sqrt{\pi} \sqrt{S} + \alpha^2 \pi \ln \left(S-\pi Q^2\right) \nonumber \\
   &-& \pi \alpha \frac{\alpha^2 + 2 Q^2}{\sqrt{\alpha^2+4 Q^2}}\, \ln\left(\frac{-\sqrt{S} + \alpha\sqrt{\pi} - \sqrt{\alpha^2 \pi+4 \pi Q^2}}{-\sqrt{S} + \alpha\sqrt{\pi} + \sqrt{\alpha^2 \pi+4 \pi Q^2}}\right) + C.
\eea
Again $C$ is an arbitrary constant.

\section{Conclusions}
\label{sec:co}

\begin{figure}[htb]
\includegraphics[angle=0,width=10cm]{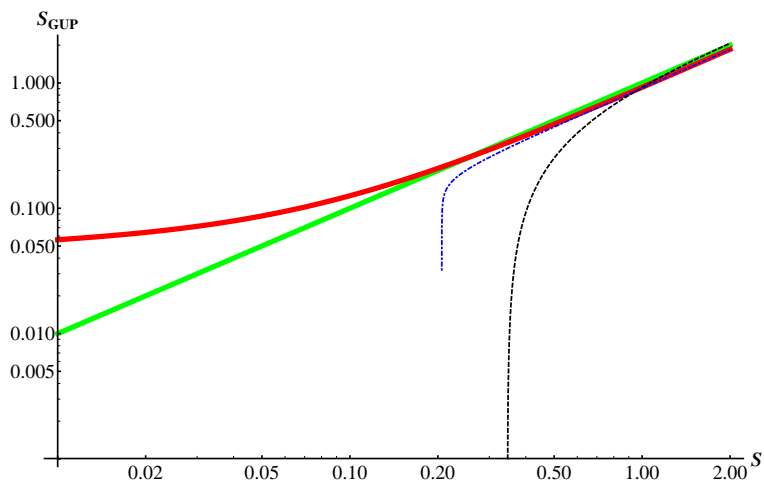}
\caption{Comparison between GUP-modifications of entropy calculated for Schwarzschild (solid curve),  Garfinkle-Horowitz-Strominger (dashed curve) and Reissner-Nordstr\"om (dash-dotted curve) black holes in log-log scale. The unmodified entropy is given by solid straight line. At very large entropy, the modifications entirely disappear. }
\label{fig:1} 
\end{figure}

\begin{figure}[htb]
\includegraphics[angle=0,width=10cm]{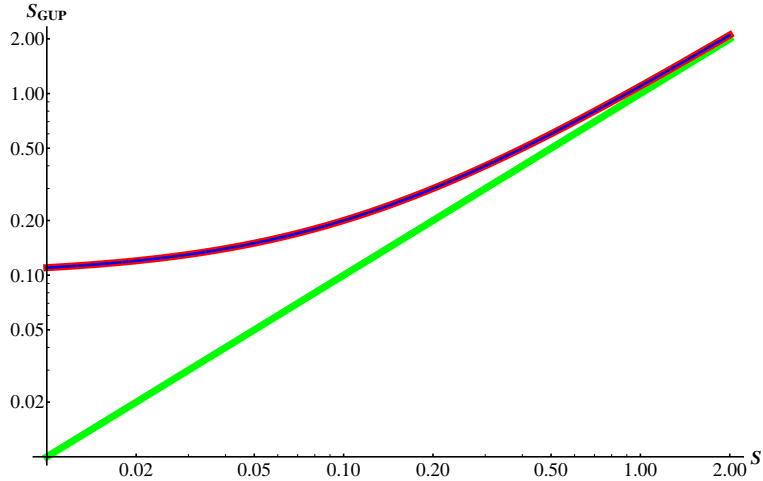}
\caption{The same as in Fig. \ref{fig:1} but at vanishing $\alpha$. }
\label{fig:2} 
\end{figure}

As discussed above, the quantum correction of the geometric entropy of charged black hole has one great advantage. Doing this one can avoid being biased in favour of a certain theory of the quantum gravity. For example, the correction to the Bekenstein-Hawking entropy, which relates the entropy to the cross-sectional area of the black hole horizon, includes a series of terms where the coefficient of the leading-order correction, the logarithmic term, is suggested as a discriminator of prospective fundamental theories for quantum gravity. It is essential to suggest a method that fixes it, but it should not depend on the utilized models for the quantum gravity. For instance, this might be the holographic principle. 

Brick wall method is used to calculate the statistical (informational) entropy of black hole. In doing this, a cutoff parameter is assuming in order to avoid the divergence near the event horizon. Because the degrees of freedom are likely dominant near the horizon, it is assumed that the brick wall method should be replaced by a thin-layer model making the calculation of entropy possible. For instance, the entropy of FLRW Universe can be given by time-dependent metric. It is found that the black hole entropy is logarithmically related to the ultraviolet regulator $\epsilon$, so that the physical entropy is limited to $\Lambda < 2 \epsilon$. 

The statistical entropy of a system likely has a clear microscopic interpretation. The question whether the black hole entropy would have a similar statistical interpretation is relevant. When comparing black hole entropy with the one that counts for the microstates $\Omega$, we can simply relate $A/4$ to $\ln \Omega$. This is valid as long as the gravity is sufficiently strong so that the horizon radius is much larger than the Compton wavelength. In order to apply GUP approach, we start with the modified momentum and statistically derive expressions for area and entropy. Then, we apply the holographic principle.  

Using GUP approach, the black hole thermodynamics and entropy get substantial corrections. The corrections are studied for three types of black holes; Schwarzschild, Garfinkle-Horowitz-Strominger and Reissner-Nordstr\"om. We found that the logarithmic divergence in the entropy-area relation turns to be positive. Furthermore we find that the entropy $S$ gets an additional terms, such as $2\, \alpha\, \sqrt{\pi}\, \sqrt{S}$, where $\alpha$ is the GUP parameter.

Figure \ref{fig:1} illustrates an intensive comparison between GUP-modifications of the entropies of Schwarzschild (solid curve),  Garfinkle-Horowitz-Strominger (dashed curve) and Reissner-Nordstr\"om (dash-dotted curve) black holes in a log-log scale. We also present the unmodified entropy by the solid straight line. We note that the modifications entirely disappear, at large values of $S$. In these calculations, the parameters, $\alpha$, $a$, $Q$ and $C$ are fixed and kept constant. It is obvious that the entropy of  Schwarzschild black hole gets a positive contribution. A considrable amount is subtracted from the entropies of  Garfinkle-Horowitz-Strominger  and Reissner-Nordstr\"om black holes. The results at vanishing $\alpha$ are illustrated in Fig. \ref{fig:2}. It is obvious that the entropies of the three types of black holes are dominated by the parameter $C$ which apparently becomes negligible at large $S$.

\appendix
 
\section{Modifications of uncertainty principle}

\subsection{Generalized (gravitational) uncertainty principle (GUP)}
\label{GUPH}

The commutator relation \cite{alii,Das:2010zf,afa2},  which are consistent with the string theory, the black holes physics and DSR leads to
 \bea
\left[\hat{x_{i}}, \hat{p_{j}} \right]=i \hbar \left[ \delta_{i j} -\alpha \left( p \delta _{i j} +\frac{p_{i} p_{j}}{p} \right)+\alpha ^{2} \left(p^{2} \delta_{ij} +3 p_{i} p_{j} \right)\right], \label{ali1}
\eea
implying a minimal length uncertainty and a maximum measurable momentum when implementing convenient representation of the commutation relations of the momentum space wave-functions \cite{amir,Tawfik:2014zca}.  The constant coefficient $\alpha=\alpha_0 \sqrt{1/(M_p\, c^2) }= \alpha_0 \sqrt{ l_p ^2/\hbar ^2}$ is  referring to the quantum-gravitational effects on the Heisenberg uncertainty principle. The momentum $\hat{p}_{j}$ and the position $\hat{x}_{i}$ operators are given as  
\bea 
\hat{x}_{i}\, \Psi (p) &=& x_{0 i}(1 -\alpha \,p_{0} +2\, \alpha^{2}\, p_{0}^{2})\, \Psi (p), \label{amm1}\nn \\
\hat{p}_{j}\, \Psi (p) &=&  p_{0 j}\, \Psi (p). \label{amm2}
\eea
We notice that $p_{0}^{2}=\sum_{j}^3p_{0 j}\,p_{0 j}$ satisfies the canonical commutation relations $\left[x_{0 i},\, p_{0 j}\right]=i\,\hbar\,\delta_{i j}$. Then, the minimal length uncertainty \cite{alii,Das:2010zf,afa2}  and maximum measurable momentum \cite{amir,Tawfik:2014zca} read
\bea 
\Delta x &\geq & (\Delta x)_{min} \approx \hbar \alpha, \nn \\
p_{max} &\approx & \frac{1}{4 \alpha},
\eea
where the maximum measurable momentum agrees with the value which was obtained in the doubly special relativity (DSR) theory \cite{Cortes:DSR2005,Tawfik:2014zca}. By using natural units, the one-dimensional uncertainty reads \cite{alii,Das:2010zf,afa2}
\bea 
\Delta x\, \Delta p &\geq & \frac{\hbar}{2} \left(1-2\, \alpha\, \Delta p +4\, \alpha^{2} \Delta p^{2}  \right). \label{ali3}
\eea
This representation of the operators product satisfies the non-commutative geometry of the spacetime \cite{amir}
\bea
\left[\hat{p_{i}},\hat{p_{j}}\right]&=&0,\nn \\
\left[\hat{x_{i}},\, \hat{x_{j}} \right] &=& - i\, \hbar\, \alpha\, \left(4\, \alpha - \frac{1}{P}\right)\, \left(1 - \alpha\, p_{0} +2 \alpha^{2}\, \vec{p_{0}}^{2}\right)\; { \hat L_{i j}}. 
\eea
The rotational symmetry does not break by the commutation relations \cite{amir}. In fact, the rotation generators can still be expressed in terms of position and momentum operators as \cite{amir,Tawfik:2014zca}
\bea
L_{i j} &=& \frac{\hat X_{i}\,  \hat P_{j} -\hat{X}_{j}\, \hat{P}_{i}}{1 -\alpha\, p_{0} +2\, \alpha^{2}\, \vec{p_{0}}^{2}}.
\eea

\subsection{Modified Dispersion Relation (MDR)}
\label{app:mdr}

Various observations support the conjectured that the Lorentz invariance might be violated. The velocity of light should differ from $c$. Any tiny adjustment leads to  modification of the energy-momentum relation and modifies he dispersion relation in vacuum state by $\delta \textit{v}$ \cite{Glashow:1a,Glashow:1b,Glashow:2,Glashow:3,Amelino98}. In particular, at the Planck scale, the modifications of energy-momentum dispersion relation have been considered in Refs. \cite{lqgDispRel1a,lqgDispRel1b,lqgDispRel2}. Two functions $p(E)$ as expansions with leading Planck-scale correction of order $L_p\, E^3$ and $L_p^2 E^4$ respectively, reads \cite{Amelino2004b},
\bea
\vec{p}^2 &\simeq & E^2 - m^2 + \alpha_1\, L_p\, E^3, \label{disprelONE} \\
\vec{p}^2 &\simeq & E^2 - m^2 + \alpha_2\, L_p^2\, E^4.
\label{disprelTWO}
\eea
These are valid a particle of mass $M$ at rest, whose position is being measured by a procedure involving a collision with a photon of energy $E$ and momentum $p$. Since the relations are originated from Heisenberg uncertainty principle for position with precision $\delta x$, one should use a photon with momentum uncertainty $\delta p \ge 1/\delta x$. Based in the argument of Ref. ~\cite{landau} in loop QG, we convert $\delta p \ge 1/\delta x$ into $\delta E \ge 1/\delta x$. By using the special-relativistic dispersion relation and $\delta E \ge 1/\delta x$, then $M \ge \delta E$. If indeed loop QG hosts a Planck-scale-modified
dispersion relation, Eq. (\ref{disprelTWO}), then$\delta p_\gamma \ge 1/\delta x$ and this required that \cite{Amelino2004b},
\begin{equation}
M  \ge \frac{1}{\delta x} \left(1 - \alpha_2 \frac{3 L_p^2}{2 (\delta x)^2}\right). \label{dis}
\end{equation}
These results apply only to the measurement of the position of a particle at rest \cite{landau}.  We can generalize these results to measurement of the position of a particle of energy $E$.
\begin{itemize}
\item In case of standard dispersion relation,  one obtains that $E  \ge 1/\delta x$ as required for a  linear dependence of entropy on area, Eq. (\ref{disprelTWO})
\item For the dispersion relation, Eq. (\ref{disprelTWO})  
\bea
E  \ge \frac{1}{\delta x} \left(1- \alpha_2 \frac{3 L_p^2}{2 (\delta x)^2}\right).
\eea
The requirements of these derivation lead in order of correction of log-area form.
\item Furthermore, 
\bea
E  \ge \frac{1}{\delta x} \left( 1+ \alpha_1 \frac{L_p}{\delta x}\right).
\eea
\end{itemize}
In case of string theory, the {\it ''reversed Bekenstein argument''} leads to  quadtraic GUP, that fits well with the string theory \cite{venegrossa,venegrossb,venegrossc} and black holes physics,
\begin{equation}
\delta x  \ge \frac{1}{\delta p} + \lambda_s^2  \delta p. \label{gup}
\end{equation}
The scale $\lambda_s$ in Eq. (\ref{gup}) is an effective string length giving the characteristic length scale which be identical with Planck length. Many researches of loop QG \cite{lqgDispRel1a,lqgDispRel1b,lqgDispRel2} support the possibility of the existence of a minimal length uncertainty and a modification in the energy-momentum dispersion relation at Planck scale.

\section{Alternative understandings of gravity}

The action of spacetime was postulated in the theory of gravitation to depend on the curvature
\bea
S(R) &=& - \frac{1}{16 \, \pi\, G} \int \sqrt{-g} \, R \, dx,
\eea
where $R$ is the invariant of the Ricci tensor. This expression refers to metrical elasticity of space or generalized forces opposing curving of space. In appendices that follow, we introduce two alternative understandings for the gravity.

\subsection{Entropic nature of gravity}

According to Eric Verlinde, it is believed that the gravitational force would have an entropic nature from holographic principle \cite{everlind1}. Thus, the gravity can be treated by thermodynamical mechanics. Introducing noncmmutative geometry implies a change in the entropy as function of the area $A$, at the surface $\Omega$ \cite{Pieroq}. The noncommutative geometry describes the microscopic structure of the quantum system. Furthermore, the GUP approach in turn implies corrections to the Newtonian law of the universal gravitational. The corrections due to linear GUP mediates another modification in the number of bits and the temperature of  black hole. 

The Newtonian law, or concretely the gravity with underlying microstructure of a quantum spacetime \cite{everlind1}, can be derived from entropy \cite{Pieroq}
\be
\Delta S_{\Omega} = k_B\, \Delta A\, \left(\frac{c^3}{4\,\hbar\, G}+\frac{\partial s(A)}{\partial A}\right), \label{genentropy}
\ee
where $s(A)$ is Bekenstein-Hawking geometric entropy. This was done under the following assumptions:
\begin{itemize}
\item in the vicinity of surface $\Omega$, the change of surface entropy becomes proportional to $\Delta x$ and the change of the radial distance becomes proportional to the mass $m$ from the surface, i.e.
\be
\Delta S_{\Omega} = 2\, \pi k_B \frac{\Delta x}{\lambda_m}.
\ee
\item The generic thermodynamic equation of state straightforwardly leads to
\be
\label{nature}
F \Delta x=T \Delta S_{\Omega}.
\ee
\item On $\Omega$, $N$ bits of information can be stored, i.e. $N=A_{\Omega}/\ell_P^2$,
where $A_\Omega$ is the area of $\Omega$ and $\ell_P$ is the Planck length.
\item The surface $\Omega$ is in thermal equilibrium at temperature $T$. Then, all bytes are equally occupied and also the energy of $\Omega$ is equipartitioned among them, i.e. $U_{\Omega}=N k_B T/2=M c^2$, where $M$ is the rest mass of the source.
\end{itemize}

By means of new uncertainty relation among coordinates, $\Delta x^\mu \Delta x^\nu\ge \theta$, where the parameter $\theta$ has the dimension of length squared and is conjectured to emerge as natural ultraviolet cutoff, the spacetime microscopic degrees of freedom are accessible \cite{Pieroq}. The commutation coordinate operators leads to $[x^\mu, x^\nu]=i \Theta^{\mu\nu}$, where $\theta=|\Theta^{\mu\nu}|$. Because of $\Omega$  uncertainty, a fundamental unit $\Delta S_{\theta}$ likely exists, i.e. $\Delta x_{min}\propto \lambda_m$. Therefore, the entropy is given as \cite{Pieroq} 
\begin{equation}
\Delta S_\Omega=\Delta S_\theta \left(\frac{\Delta x}{\Delta x_{min}}\right), 
\end{equation}
where $\Delta x_{min}=\alpha^2\,\lambda_m /(8\, \pi)$. This is the fundamental surface, which coincides with $\theta$ and leads to $N=A_\Omega/\theta$. 

On the other hand, the Planck scale and $\alpha$-parameter  in noncommutative geometry lead to corrections to entropy and temperature, respectively \cite{Pieroq},
\begin{eqnarray}
\Delta S_{\theta} &=& k_B \theta \left(\frac{c^3}{4 \hbar G}+\frac{\partial s}{\partial A}\right), \qquad\qquad T = \frac{M}{r^2}\frac{\theta\, c^2}{2\, \pi\, k_B}. \label{temp2}
\end{eqnarray}
From Eq. (\ref{nature}), the Newtonian law of universal gravity gets a positive correction, i.e. derivative of entropy to area gives a positive value  \cite{Pieroq}.   
\begin{equation}
F=\frac{Mm}{r^2}\left(\frac{4 c^3 \theta^2}{\hbar\alpha^2}\right)\left[\frac{c^3}{4\hbar G}+\frac{\partial s(A)}{\partial A}\right]. \label{ncforce}
\end{equation}
It is obvious that the first term of Eq. (\ref{ncforce}) is nothing but the Newtonian law, if $\theta=\alpha\ell_P^2$,
\begin{equation}
F = \frac{GMm}{r^2}\left[1+4\ell_P^2\frac{\partial s}{\partial A}\right]. \label{Newtoncorr}
\end{equation}

\subsection{Planck-Kleinert world crystal}

The world crystal model \cite{wcm1,wcm3} was introduced as an alternative understanding for the gravity. In this model, the fact that crystals with defects should have the same non-Euclidean geometry as spaces with curvature and torsion was exploited. Thus the world crystal is believed to represent a model for emergent or induced gravity \cite{everlind1} in an Einstein-Cartan theory of gravitation, which embraces GR.  The induced (emergent) gravity was originally proposed by Sakharov in 1967 \cite{sakharf67}. This was based on the idea that in QG that space-time background emerges as a mean field approximation of underlying microscopic degrees of freedom, similar to the fluid mechanics approximation of Bose-Einstein condensates. 

In the Planck-Kleinert crystal, the geometry of Einstein and Einstein-Cartan spaces can be considered as being a manifestation of the defect structure of a crystal whose lattice spacing is of the order of the Planck length. This is an ideal face-centred cubic crystal showing the Frenkel disorder and mechanical properties independent on strain and consisting of Planck particles with conserved mass, momentum and energy \cite{wcm3}. The equation of internal energy conservation has the consequence of  existing waves involving temperature, but not the mechanical potential variations, i.e. the second sound proposed by Landau and Lifschitz.  At the Planck length, each particle exerts a short range force. Due to mechanical energy conservation, the gravity field generated by the immobile massive body in the crystal can be estimated. The immobile body implies a quasistationary situation and the deformation, mass and energy in a space occupied by the body are fixed. The diffusing Planck particles are the initiator of gravity. This is the interpretation of gravity, i.e. curvature is due to rotational defects and torsion due to translational defects. Even, the classical Newton’s law of gravity can be derived from the world crystal model, which - in turn - illustrates that the world may have, at Planck distances, quite different properties from those predicted by string theorists. The matter creates defects in spacetime which generates curvature and all the effects of GR. 

The world crystal model found various applications including GUP \cite{wcmB1}. In the particular case, when energies lie near the border of the Brillouin zone, i.e. for Planckian energies, the uncertainty relation for position and momenta does not  pose any lower bound on involved uncertainties.
For Planckian lattices, GUP reads
\bea
\Delta X_{\epsilon}\, \Delta P_{\epsilon} \geq \frac{\hbar}{2} \left(1- \frac{\epsilon^2}{2 \hbar^2} \left(\Delta P_{\epsilon}\right)^2\right),
\eea
where $\epsilon$ has the order of Planck length. Hence the world-crystal Universe can become deterministic at Planckian energies. It was concluded in Ref. \cite{wcmB1} that the lattice uncertainty relations seem to resemble the quadratic GUP approach.





\begin{thebibliography}{99}

\bibitem{hist1a} J. D. Bekenstein, Phys. Rev. D {\bf 7}, 2333 (1973); Phys. Rev. D {\bf 9}, 3292 (1974).
\bibitem{hist1b} S. W. Hawking, Nature {\bf 248}, 30 (1974).
\bibitem{hist1c} S. W. Hawking, Commun. Math. Phys. {\bf 43}, 199 (1975).

\bibitem{tsallis} C. Tsallis, private communication

\bibitem{refff2} A.J.M. Medved and E.C. Vagenas, Phys. Rev. D {\bf 70}, 124021 (2004).

\bibitem{Tawfik:2015pqa} Abdel Nasser Tawfik, and Abdel Magied Diab, {\it ''Black Hole Corrections due to Minimal Length and Modified Dispersion Relation''}, to appear in Int. J. Mod. Phys. A, 1502.04562 [gr-qc].

\bibitem{Wotae} Huyunjoo Lee, Sung-Won Kim and Won T. Kim, Phys. Rev. D {\bf 54}, 6559, (1996). 

\bibitem{hist4} A. Strominger and C. Vafa, Phys. Lett. B {\bf 379}, 99 (1996).

\bibitem{hist5} A. Ashtekar, J. Baez, A. Corichi and K. Krasnov, Phys. Rev. Lett. {\bf 80}, 904 (1998).

\bibitem{sAclass} A. Gould, Phys. Rev. D {\bf 35}, 449 (1987). 

\bibitem{hist7} G. Gour and A.E. Mayo, Phys. Rev. D {\bf 63}, 064005 (2001).

\bibitem{hist8} M.R. Setare, Eur. Phys. J. C {\bf 33}, 555 (2004).

\bibitem{hist9} R.K. Kaul and P. Majumdar, Phys. Rev. Lett. {\bf 84}, 5255 (2000).

\bibitem{hist16} R. Bousso, Rev. Mod. Phys. {\bf 74}, 825 (2002).

\bibitem{everlind1} E.~P.~Verlinde,  JHEP {\bf 1104}, 029 (2011). 

\bibitem{entr5} G. t'Hooft, Nucl. Phys. B {\bf 256}, 727 (1985).

\bibitem{entr6a} L. Susskind and J. Uglum, Phys. Rev. D {\bf 50}, 2700 (1994).

\bibitem{entr6b} T. Jacobson, Phys. Rev. D {\bf 50}, 6031 (1994).

\bibitem{entr6c} S. P. de Alwis and N. Ohta, Phys. Rev. D {\bf 52}, 3529 (1995)].

\bibitem{entr6d} J. G. Demers, R. Lafrance and R. C. Myers, Phys. Rev. D {\bf 52}, 2245 (1995).

\bibitem{entr6e} S. Mukohyama, Phys. Rev. D {\bf 61}, 124021 (2000).

\bibitem{entr6f} S. W. Kim, W. T. Kim, Y. J. Park and H. Shin, Phys. Lett. B {\bf 392}, 311 (1997).

\bibitem{entr6g}  A. Ghosh and P. Mitra, Phys. Rev. Lett. {\bf 73}, 2521 (1994).

\bibitem{entr6h} J. Ho, W. T. Kim, Y. J. Park and H. Shin, Class. Quant. Grav. {\bf 14}, 2617 (1997).

\bibitem{sabine} S. Hossenfelder,  Living Rev. Rel. {\bf 16}, 2 (2013). 



\bibitem{alii} A. Ali, S. Das and E. C. Vagesas, Phys. Lett. B {\bf 678}, (2009).

\bibitem{reffff1} W. Kim, Y.-W. Kim, and Y.-J. Park, Phys. Rev. D {\bf 75}, 127501 (2007).  

\bibitem{Tawfik:2014zca} A. Tawfik and A. Diab, {\it ''Generalized Uncertainty Principle: Approaches and Applications''},  Int. J. Mod. Phys. D {\bf 23} 1430025 (2014).

\bibitem{gupp1} L.N. Chang, D. Minic, N. O. Karuma and T. Takeachi, Phys. Rev. D {\bf 65}, 125028 (2002).


\bibitem{refff4} D. Garfinkle, G. T. Horowitz, A. Strominger, Phys. Rev. D {\bf 43}, 3140 (1991), Erratum-ibid. D {\bf 45}, 3888 (1992). 

\bibitem{RN1} H. Reissner, 
Annalen der Physik {\bf 50}, 106-120 (1916).

\bibitem{RN2} G. Nordstr\"om, {\it ''On the Energy of the Gravitational Field in Einstein's Theory."} Proc. Kon. Ned. Akad. Wet. 20, 1238-1245, (1918).

\bibitem{refff5} Zhao Hai-Xia, Li Huai-Fan, Hu Shuang-Qi and Zhao Ren, Commun. Theor. Phys. {\bf 48}, 465 (2007).







\bibitem{Das:2010zf} S.~Das, E.~C.~Vagenas and A.~F.~Ali, 
Phys. Lett.  B {\bf 690}, 407 (2010). 

\bibitem{afa2}  A. Farag Ali, S. Das and E. C. Vagenas, 
Phys. Rev. D {\bf 84}, 044013 (2011). 


    
\bibitem{amir} K. Nozari and A. Etemadi, 
Phys. Rev. D {\bf 85}, 104029 (2012). 


\bibitem{Tawfik:2013uza} A. Tawfik, 
JCAP {\bf 1307}, 040 (2013). 
                        
\bibitem{Ali:2013ma} A. F. Ali and A. Tawfik, 
Adv. High Energy Phys. {\bf 2013}, 126528 (2013). 

\bibitem{Ali:2013ii} A. F. Ali and A. Tawfik, 
Int. J. Mod. Phys. D {\bf 22}, 1350020 (2013). 

\bibitem{Tawfik:2012he} A. Tawfik, H. Magdy and A.Farag Ali, 
Gen. Rel. Grav. {\bf 45}, 1227-1246 (2013). 

\bibitem{Tawfik:2012hz} A. Tawfik, H. Magdy and A.F. Ali, {\it ''Lorentz Invariance Violation and Generalized Uncertainty Principle''}, 1205.5998 [physics.gen-ph].

\bibitem{Elmashad:2012mq} I. Elmashad, A.F. Ali, L.I. Abou-Salem, Jameel-Un Nabi and A. Tawfik, 
Trans. Theor. Phys, {\bf 1}, 106 (2014). 


\bibitem{dinverno} Roy D'Inverno, {\it ''Introducing Einstein's Relativity''}, (Clarendon Press, Oxford, 1992). 

\bibitem{Dirac} P.A.M. Dirac, {\it "The Principles of Quantum Mechanics"}, (Clarendon Press, Oxford, 1958).


\bibitem{Cortes:DSR2005} J. Cortes and J. Gamboa, 
Phys. Rev. D {\bf 71}, 065015 (2005). 


 

\bibitem{Wesson:2004} P. S. Wesson, Mod. Phys. Lett. A {\bf 19}, 1995 (2004).

\bibitem{20} Michele Maggiore, 
Phys. Rev. D {\bf 49}, 5182-5187 (1994). 

\bibitem{12} J. Cortes and J. Gamboa, 
Phys. Rev. D {\bf 71}, 065015 (2005). 

    
\bibitem{nature2012} I. Pikovski,	 M. R. Vanner,	 M. Aspelmeyer,	 M. S. Kim	and C. Brukner, 
Nature Phys. {\bf 8}, 393-397 (2012).

\bibitem{sdual} Luis A. Anchordoqui,V. Barger, H. Goldberg, Xing Huang and D. Marfatia, Phys. Lett. B {\bf 734}, 134-136 (2014). 


\bibitem{JSchwarz:2002}S. M. Leach, A. R. Liddle, J. Martin and D. J. Schwarz,
Phys. Rev. D {\bf 66}, 023515 (2002). 

\bibitem{Montonen:1977} C. Montonen and D. I. Olive, Phys. Lett. B {\bf 72}, 117 (1977).

\bibitem{Starobinsky} Alexei A. Starobinsky, Phys. Lett. B {\bf 91}, 99-102 (1980).
 
 \bibitem{Amelino2004b} 
 G. Amelino-Camelia, M. Arzano and A. Procaccini, 
 Int. J. Mod. Phys. D {\bf 13}, 2337 (2004). 
 
\bibitem{Liddle:2003} A. R. Liddle, {\it ''Introduction to modern cosmology''} (Wiley, Weinheim, 2003).

\bibitem{Linde:2002} A. D. Linde, (ed. S. Bonometto, V. Gorini and U. Moschella) {\it Inflationary cosmology and creation of matter in the Universe. In Modern cosmology}, (Institute of Physics Publishing, Bristol, 2002).

\bibitem{Liddle:1993} Andrew R. Liddle, David H. Lyth, Phys. Rept. {\bf 231}, 1-105 (1993).



\bibitem{sanchez-2007} J. C. B. Sanchez, K. Dimopoulos and D. H. Lyth, JCAP {\bf 0701}, 015 (2007).

\bibitem{3} M. Maggiore, 
Phys. Lett. B {\bf 304}, 65 (1993). 

\bibitem{Scardigli} F. Scardigli, 
Phys. Lett. B {\bf 452}, 39 (1999). 

\bibitem{7} F. Scardigli and R. Casadio, 
Int. J. Mod. Phys. D {\bf 18}, 319-327 (2009). 

\bibitem{lqgDispRel1a} R.~Gambini and J.~Pullin, 
Phys.~Rev.~D {\bf 59}, 124021 (1999). 
\bibitem{lqgDispRel1b} J.~Alfaro, H.A.~Morales-Tecotl and L.F.~Urrutia,
Phys.~Rev.~Lett.~{\bf 84}, 2318-2321 (2000). 

\bibitem{lqgDispRel2}  G.~Amelino-Camelia, L.~Smolin and A.~Starodubtsev, 
Class. Quant. Grav. {\bf 21}, 3095-3110 (2004). 

    
\bibitem{nature2012} I. Pikovski,	 M. R. Vanner,	 M. Aspelmeyer, M. S. Kim	and C. Brukner, 
Nature Phys. {\bf 8}, 393-397 (2012). 

\bibitem{venegrossa} G. Veneziano, Europhys.~Lett.~{\bf 2}, 199 (1986).
\bibitem{venegrossb} D.J. Gross and P.F. Mende, Nucl.~Phys.~B {\bf 303}, 407 (1988). 
\bibitem{venegrossc} D. Amati, M. Ciafaloni and G. Veneziano, Phys.~Lett.~B {\bf 216}, 41 (1989).

\bibitem{landau} E.~M.~Lifshitz, L.~P.~Pitaevskii and
V.~B.~Berestetskii, ``Landau-Lifshitz Course of Theoretical Physics, Volume 4: Quantum Electrodynamics"
(Reed Educational and Professional Publishing, 1982).


\bibitem{Glashow:1a} 
S. Coleman and S. L. Glashow, 
Phys. Rev. D {\bf 59}, 116008 (1999). 

\bibitem{Glashow:1b} D. Colladay and V. A. Kostelecky, 
Phys. Rev. D {\bf 58}, 116002 (1998). 

\bibitem{Glashow:2} 
F.W. Stecker and Sheldon L. Glashow, 
Astropart. Phys. {\bf 16} 97-99  (2001). 

\bibitem{Glashow:3} 
S. R. Coleman and S. L. Glashow, 
Phys. Lett. B {\bf 405}, 249-252 (1997). 

\bibitem{Amelino98} 
G. Amelino-Camelia, J. Ellis, N. F. Mavromatos, D. V. Nanopoulos, S. Sarkar, 
Nature {\bf 393}, 763 (1998). 





\bibitem{Pieroq} Piero Nicolini, 
Phys. Rev. D {\bf 82}, 044030 (2010). 

\bibitem{wcm1}
H. Kleinert, 
Annalen der Physik {\bf 44}, 117  (1987).

\bibitem{wcm3} M. Danielewski, 
Zeitschrift f\"ur Naturforschung A {\bf 62}, 56 (2007).


\bibitem{sakharf67} A. D. Sakharov, Doklady Akademii Nauk SSSR,  {\bf 177},  70-77 (1967) translated into English Gen. Rel. Grav. {\bf 32}, 365-367 (2000).

\bibitem{wcmB1} Petr Jizba, Hagen Kleinert, and Fabio Scardigli, 
Phys. Rev. D {\bf 81}, 084030 (2010) .



\end{thebibliography}
\end{document}